\documentclass[prl,letterpaper,twocolumn,preprintnumbers,nofootinbib]{revtex4}
\usepackage[a4paper, hdivide={1.91cm,,1.165cm}, vdivide={1.83cm,,2.8cm}]{geometry}
\usepackage{slashed}
\usepackage{amsmath,amssymb}
\usepackage{graphicx}
\usepackage{units}
\usepackage{xcolor}
\usepackage[hyperfootnotes=false,colorlinks,citecolor=blue]{hyperref}

\newcommand{\beq}{\begin{equation}}
\newcommand{\eeq}{\end{equation}}
\newcommand{\bea}{\begin{eqnarray}}
\newcommand{\ena}{\end{eqnarray}}

\def \L {\mathcal{L}} %Lagrangian density

\def \epsilon {\varepsilon} %different epsilon symbol

\newcommand{\hc}{\ensuremath{\text{h.c.}}}
\newcommand{\BR}{\ensuremath{\text{BR}}}
\newcommand{\ii}{\ensuremath{\text{i}}}
\newcommand{\tr}{\ensuremath{\text{Tr}}}
\newcommand{\re}{\ensuremath{\text{Re}}}
\newcommand{\im}{\ensuremath{\text{Im}}}

\allowdisplaybreaks
\begin{document}
%%%%%%%%%%%%%%%%%%%%%%%%%%%%%%%%%

\title{
$W$-boson mass in the triplet seesaw model
}

\author{Julian Heeck}
\email{heeck@virginia.edu}
\affiliation{Department of Physics, University of Virginia,
Charlottesville, Virginia 22904-4714, USA}

%%%%%%%%%%%%%%%%%%%%%%%%%%%%%%%%%%
%%%%%%%%%%%%%%%%%%%%%%%%%%%%%%%%%%

\begin{abstract}
The CDF collaboration has recently published a precision measurement of the $W$-boson mass that differs from the Standard Model prediction by seven standard deviations. This result can be explained with additional electroweak multiplets that either break the custodial symmetry or contribute to oblique parameters at loop level. Here, we study one of the best-motivated scenarios involving new multiplets: the type-II seesaw model, which involves a scalar triplet that generates Majorana neutrino masses and can furthermore resolve the $W$-boson mass discrepancy. This favors a doubly-charged scalar with mass between 100 and \unit[200]{GeV} as well as other scalars with a fixed mass splitting. The entire preferred parameter space is testable at the LHC.
\end{abstract}

%%%%%%%%%%%%%%%%%%%%%%%%%%%%%%%%%
%%%%%%%%%%%%%%%%%%%%%%%%%%%%%%%%%
\maketitle

%%%%%%%%%%%

The CDF collaboration at Fermilab  has recently published a new precision measurement of the $W$-boson mass using their full $\unit[8.8]{fb^{-1}}$ data set~\cite{CDF:2022hxs},
\begin{align}
m_W^\textrm{CDF}= \unit[(80.4335 \pm 0.0094)]{GeV},
\label{eq:CDF}
\end{align}
which deviates by $7\sigma$ from the Standard Model (SM) prediction $m_W^{\rm SM} = \unit[(80.357 \pm 0.004)]{GeV}$~\cite{Awramik:2003rn}.
This deviation has drastic consequences for particle physics since it implies a breakdown of the cherished custodial-symmetry relation $m_W = m_Z \cos\theta_W$ between the  independently measured weak mixing angle $\theta_W$ and the $W$ and $Z$-boson masses.

Simple possible explanations of the CDF result can be found at tree level in models where the electroweak symmetry is broken by $SU(2)_L$ multiplets larger than a doublet~\cite{Strumia:2022qkt,Asadi:2022xiy,Bagnaschi:2022whn,DiLuzio:2022xns,Du:2022brr,Perez:2022uil,Ghoshal:2022vzo,Borah:2022obi,Athron:2022isz,Batra:2022org}, or  in models with non-degenerate electroweak multiplets that break custodial symmetry at loop level~\cite{Fan:2022dck,Zhu:2022tpr,Athron:2022qpo,Strumia:2022qkt,Liu:2022jdq,Asadi:2022xiy,Song:2022xts,Bahl:2022xzi,Babu:2022pdn,Biekotter:2022abc,Crivellin:2022fdf,Heo:2022dey,Ahn:2022xeq,Kawamura:2022uft,Nagao:2022oin,Carpenter:2022oyg,Popov:2022ldh,Arcadi:2022dmt,Chowdhury:2022moc,Ghorbani:2022vtv,Bhaskar:2022vgk,Lee:2022nqz,Baek:2022agi,Lu:2022bgw,Han:2022juu,Heckman:2022the,Cao:2022mif}.
In this letter, we discuss a well known simple model that has both of these properties and, importantly, also solves the neutrino mass problem of the SM: the triplet (or type-II) seesaw model~\cite{Konetschny:1977bn,Magg:1980ut,Schechter:1980gr,Cheng:1980qt,Mohapatra:1980yp}.
%\footnote{This model was already discussed in this context in Refs.~\cite{Cheng:2022jyi,Kanemura:2022ahw}, but we disagree with their conclusions.}

We extend the SM by an $SU(2)_L$ triplet with hypercharge $+2$, conveniently written in matrix form as
\begin{align}
\Delta = \begin{pmatrix}
\Delta^+/\sqrt{2} & \Delta^{++}\\
\Delta^0 & -\Delta^+/\sqrt{2}
\end{pmatrix} .
\end{align}
Its only renormalizable Yukawa coupling is to the left-handed lepton doublets $L_{e,\mu,\tau}$,
\begin{align}
\L_\text{Yukawa} =  \sum_{\alpha,\beta = e,\mu,\tau}Y_{\alpha\beta} \bar{L}^c_\alpha \ii\tau_2 \Delta L_\beta + \hc ,
\end{align}
which induces a Majorana neutrino mass matrix $M_\nu = \sqrt{2} Y v_\Delta$ once the neutral component $\Delta^0$ acquires a vacuum expectation value (VEV) $\langle \Delta^0\rangle = v_\Delta/\sqrt{2}$. This immediately solves the biggest problem of the SM -- its massless neutrinos -- and furthermore fixes the flavor structure of the Yukawa couplings $Y$ in terms of neutrino masses and mixing angles, most of which are precisely measured already~\cite{deSalas:2020pgw}. Despite accurate measurements of the neutrino mass splittings through neutrino oscillations, the \emph{absolute} neutrino mass scale is not known yet, with upper limits coming from beta-decay measurements, $m_\nu <\unit[0.8]{eV}$~\cite{KATRIN:2021uub}, and cosmology, $\sum m_\nu < \unit[0.12]{eV}$~\cite{Planck:2018vyg}. 
Notice that the overall normalization of $Y$ remains completely uncertain even with full knowledge of $M_\nu$, and could range from  $\mathcal{O}(1)$ if $v_\Delta \sim \unit{eV}$ to $\mathcal{O}(10^{-9})$ if $v_\Delta \sim \unit{GeV}$.

In addition to the Yukawa couplings, $\Delta$ has gauge interactions specified by its gauge group representation as well as couplings to the SM Higgs doublet $\Phi$ in the scalar potential $V$:
\begin{align}
V &= -m_\Phi^2 \Phi^\dagger\Phi + \frac{\lambda}{4}(\Phi^\dagger\Phi )^2 +  \lambda_1 \Phi^\dagger\Phi \,\tr [\Delta^\dagger \Delta] \nonumber\\
&\quad + \lambda_2 (\tr[\Delta^\dagger \Delta])^2+ \lambda_3 \tr[\Delta^\dagger \Delta\Delta^\dagger \Delta]+ \lambda_4 \Phi^\dagger \Delta \Delta^\dagger \Phi  \nonumber\\
&\quad+
\tilde M_\Delta^2 \tr [\Delta^\dagger \Delta] + \left[\mu \Phi^T \ii \tau_2\Delta^\dagger\Phi +\hc\right] .
\end{align}
Minimization yields the triplet VEV $v_\Delta \simeq \mu v^2/(\sqrt{2} \tilde M_\Delta^2)$, which we will assume to be much smaller than the Higgs doublet VEV $v= \sqrt{2}\langle \Phi\rangle \simeq \unit[246]{GeV}$ to be consistent with data discussed below.
In this limit of $v_\Delta \ll v$, the mass eigenstates that dominantly come from the triplet, $H^{++}\simeq \Delta^{++}$, $H^+\simeq \Delta^{+}$, $H\simeq\sqrt2 \re\,\Delta^0$, and $A\simeq \sqrt2\im\,\Delta^0$, have mass splittings
\begin{align}
m_H^2 \simeq m_A^2 \simeq m_{H^+}^2+\frac{\lambda_4 v^2}{4}\simeq m_{H^{++}}^2+\frac{\lambda_4 v^2}{2} \,,
\label{eq:splittings}
\end{align}
specified exclusively by the coupling $\lambda_4$~\cite{Mandal:2022zmy}.

In this setup, we can now calculate the custodial symmetry breaking or the $W$-boson mass. Since a contribution to $m_W$ can easily modify other observables connected to electroweak precision data, it behooves us to perform a consistent fit to all data, including the new CDF result. In addition to the SM, this fit needs to include higher-dimensional operators to describe the new physics contributions, which in our case correspond to the well-established oblique parameters $S$, $T$, and
$U$~\cite{Peskin:1990zt,Peskin:1991sw}, the latter typically being negligible. 
The oblique contributions to $m_W$ for $U=0$ take the general form~\cite{Maksymyk:1993zm}
\begin{align}
m_W \simeq m_W^\text{SM}\left[ 1- \frac{\alpha_\text{EM}}{4(1-2 s_W^2)}\left( S - 2 (1-s_W^2) T\right)\right] ,
\end{align}
with $s_W^2=\sin^2\theta_W\simeq 0.23$. The CDF result then requires non-zero oblique parameters around $T\simeq 0.18 + 0.65 S$. The host of additional electroweak data provides orthogonal constraints that restrict $S$ and $T$ towards small positive values.
A dedicated fit has been performed recently in Ref.~\cite{Asadi:2022xiy}, quoting best fit values $(S,T) = (0.17,0.27)$ that deviate from the SM-predicted $(0,0)$ due to the pull from the CDF result. We will employ these fit results from Ref.~\cite{Asadi:2022xiy} in the following and map them onto parameters of the triplet model.

\begin{figure}[tb]
    \centering
    \includegraphics[width=0.38\textwidth]{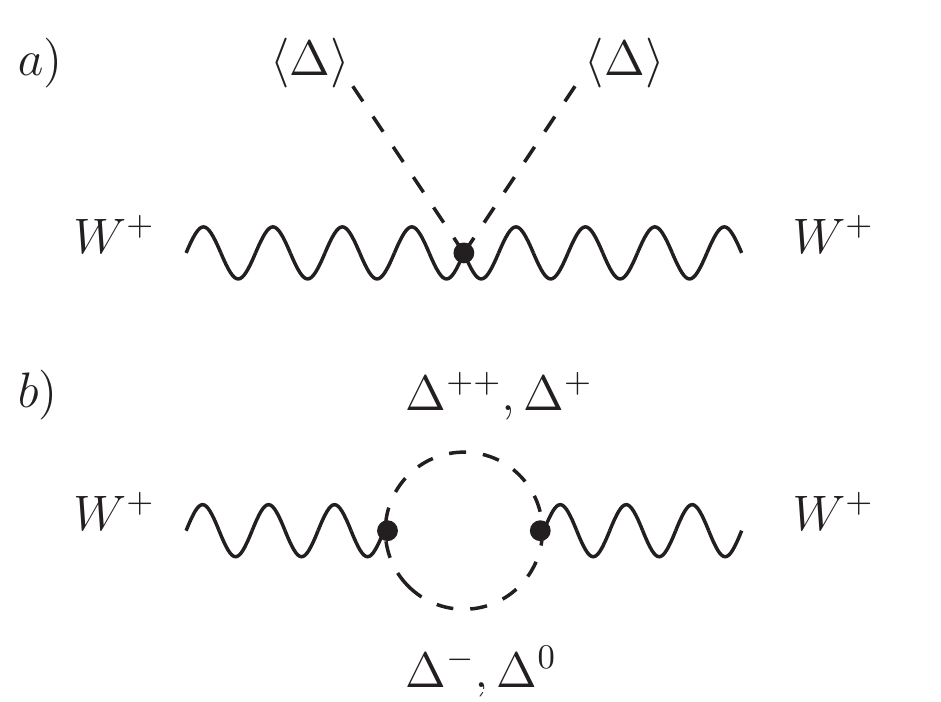}
    \caption{Example Feynman diagrams for the a) tree-level contribution of the triplet $\Delta$ to the $W$-boson mass from the triplet vacuum expectation value; b) loop-level contribution induced by non-degenerate triplet components. Similar diagrams exist for the $Z$ boson.    
     }
    \label{fig:diagrams}
\end{figure}

At tree-level, the triplet seesaw model contributes only to $T$, usually expressed through the $\rho$ parameter~\cite{Ross:1975fq} via $\rho = 1+\alpha_\text{EM} T$,
\begin{align}
T \simeq -\frac{2}{\alpha_\text{EM}}\, \frac{v_\Delta^2}{v^2} \simeq - 4\times 10^{-3} \left(\frac{v_\Delta}{\unit{GeV}}\right)^2,
\label{eq:tree_level}
\end{align}
illustrated in Fig.~\ref{fig:diagrams} (a)~\cite{Schechter:1980gr}.
This has the wrong sign, making the tension with the CDF result even worse. Luckily, at one-loop level (Fig.~\ref{fig:diagrams} (b)) we can obtain positive contributions to $T$ that dominate over the tree-level contribution, as long as we keep the latter small by demanding $v_\Delta \lesssim \unit[1]{GeV}$.
The full expressions for $S$, $T$, and $U$ are given in Refs.~\cite{Lavoura:1993nq,Chun:2012jw} and are used in our numerical analysis, but let us briefly consider the approximate limit of tiny $\lambda_4$ and heavy triplet masses, which gives~\cite{Mandal:2022zmy}
\begin{align}
S &\simeq - \frac{(2-4 s_W^2+5 s_W^4) m_Z^2}{30\pi m_{H}^2}+\lambda_4 \frac{v^2}{6\pi m_{H}^2} \,,\\
&\simeq  3\times 10^{-3}\frac{\lambda_4 - 0.04}{(m_H/\unit{TeV})^2} \,,\nonumber\\
T &\simeq  \frac{ v^2 \lambda_4^2}{192\pi^2 \alpha_\text{EM}  m_{H}^2} \simeq \frac{4\times 10^{-3}}{(m_H/\unit{TeV})^2}\,\lambda_4^2\,, \\
U&\simeq  \frac{(2-4 s_W^2 +5 s_W^4) m_Z^2 - 2 m_W^2}{30\pi m_{H}^2} \simeq \frac{-2\times 10^{-5}}{(m_H/\unit{TeV})^2}\,.
\end{align}
These expressions already illustrate several important points: i) the $U$ parameter is suppressed compared to $S$; ii) the $T$ parameter is positive at one loop and hence contributes positively to the $W$ mass, as required by CDF; iii) to keep $S$ positive as well, as required for the fit, we need $\lambda_4 > 0$. Albeit obvious, it bears emphasizing that the oblique parameters vanish when the triplet is pushed to very high masses, in accordance with the decoupling theorem, which will in turn provide \emph{upper} bounds on the new masses when the CDF result is to be explained.

\begin{figure}[tbh]
    \centering
    \includegraphics[width=0.41\textwidth]{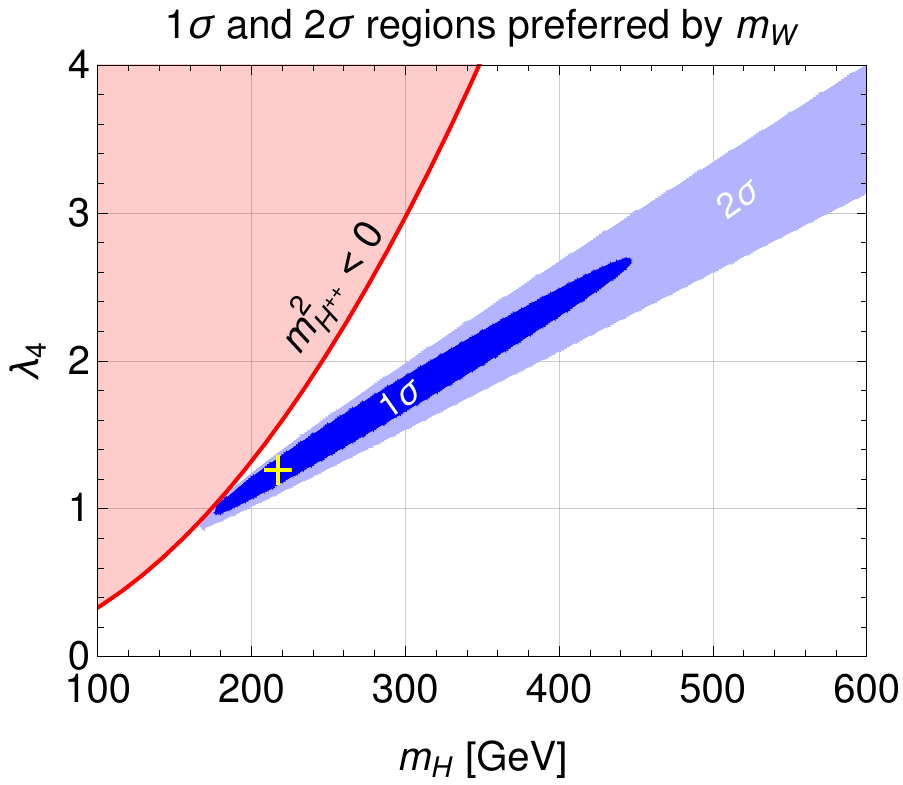}\\[2ex]
    \includegraphics[width=0.43\textwidth]{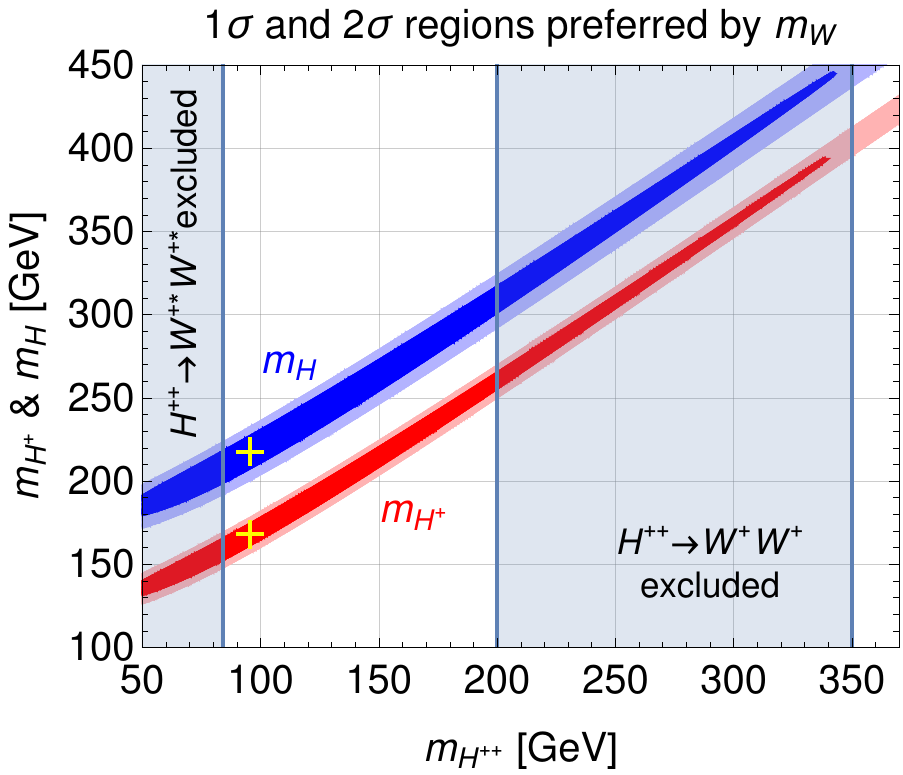}\\[2ex]
    \includegraphics[width=0.43\textwidth]{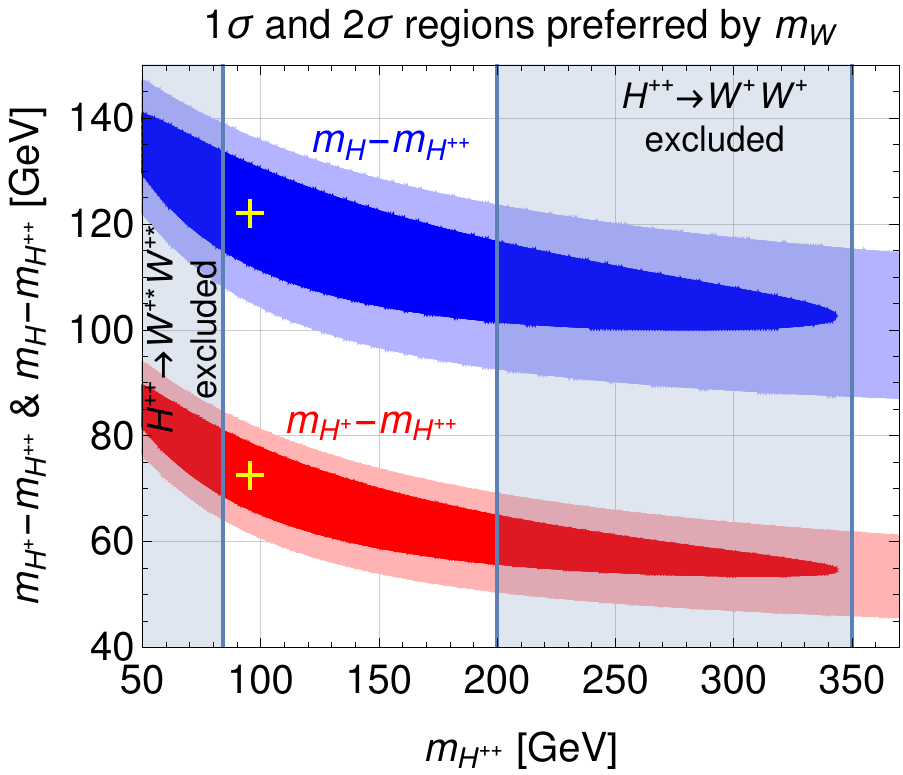}
    \caption{$1\sigma$ and $2\sigma$ preferred regions for $m_W$ in the triplet seesaw model, using the fit from Ref.~\cite{Asadi:2022xiy}. Yellow markers denote the best-fit points. See text for details.
     }
    \label{fig:ST}
\end{figure}

Using the full expressions for the oblique parameters~\cite{Lavoura:1993nq,Chun:2012jw} and neglecting the tree-level contribution from Eq.~\eqref{eq:tree_level} by making $v_\Delta \ll v$, we find the required values in the $\lambda_4$--$m_H$ plane that resolve the $m_W$ anomaly in Fig.~\ref{fig:ST} (top). Positive values $\lambda_4\sim 1$ are  required for the desired mass splitting of the multiplet components, which is well within the perturbative regime~\cite{Mandal:2022zmy} and predicts the hiearchy $m_{H^{++}} < m_{H^+}< m_{H}\simeq m_A$.\footnote{Negative values for $S$ and $\lambda_4$  are only allowed at $2\sigma$ in this fit~\cite{Asadi:2022xiy} and require rather large $|\lambda_4|\gtrsim 5$; see Ref.~\cite{Kanemura:2022ahw} for a discussion of the triplet model in this parameter region.} 
The neutral scalar $H$ has to lie between $200$ and $\unit[450]{GeV}$ to accommodate the CDF measurement within $1\sigma$ and the common mass-squared difference $m_H^2 - m_{H^+}^2 = m_{H^+}^2  - m_{H^{++}}^2 $ is between $(\unit[120]{GeV})^2$ for small $m_H$ and $(\unit[200]{GeV})^2$ for large $m_H$, respectively.
Trading the coupling $\lambda_4$ for the mass splittings yields Fig.~\ref{fig:ST} (bottom). 
%From $\lambda_4 >0$ we predict the hiearchy $m_{H^{++}} < m_{H^+}< m_{H}\simeq m_A$ and a mass splitting around 60 to $\unit[80]{GeV}$ between the components.
Notice that in the region of interest, $U/S\simeq 0.1$--$0.3$, which should be sufficiently small to indeed neglect $U$ in the global fit, although $U$ is larger than naively expected for this light triplet.

We have chosen $v_\Delta\ll v$ for several reasons: 1) to obtain small neutrino masses without tiny Yukawa couplings; 2) to obtain correlated and predictive mass splittings in Eq.~\eqref{eq:splittings}; 3) to be able to encode the triplet effects in the three Peskin--Takeuchi parameters $S$, $T$, and $U$~\cite{Peskin:1991sw,Lavoura:1993nq}.
A large $v_\Delta$ furthermore requires finetuning between the tree-level and one-loop contribution to $T$ that pushes $\lambda_4$ -- and by extension $m_H$ -- to even larger values. We therefore consider $v_\Delta\ll v$ the most natural region of parameter space. Values for $v_\Delta$ above GeV require a different renormalization scheme, as discussed in Refs.~\cite{Blank:1997qa,Aoki:2012jj}, but still allow for a resolution of CDF's $m_W$ result~\cite{Kanemura:2022ahw}.

The electroweak fit, including Eq.~\eqref{eq:CDF}, fixes the allowed range of masses for the triplet scalars, but additional constraints arise from lepton flavor violation and collider searches. At $1\sigma$, the triplet masses are required to be below $\unit[450]{GeV}$, even $\unit[350]{GeV}$ for the doubly charged scalar $H^{++}$. These scalars contribute to  lepton-flavor-violating decays, notably~\cite{Pich:1984uoh,Mandal:2022zmy}
\begin{align}
&\BR (\mu\to e\gamma)\simeq \frac{\alpha_\text{EM} \left|(M_\nu^\dagger M_\nu)_{e\mu}\right|^2}{48\pi G_F^2 v_\Delta^4}\left(\frac{1}{m_{H^{+}}^2}+\frac{8}{m_{H^{++}}^2}\right)^2,\\
&\BR (\mu^+\to e^+e^-e^+)\simeq 4\frac{\left|(M_\nu)_{ee} (M_\nu)_{\mu e}\right|^2}{G_F^2 v_\Delta^4 m_{H^{++}}^4}\,.
\end{align}
The combination $\left|(M_\nu^\dagger M_\nu)_{e\mu}\right|^2$ only depends on known neutrino oscillation parameters~\cite{Chakrabortty:2012vp} and is limited from below by $(\unit[0.016]{eV})^4$, using the $2\sigma$ range from Ref.~\cite{Esteban:2020cvm}. The current limit $\BR (\mu\to e\gamma)<4.2\times 10^{-13}$~\cite{MEG:2016leq} then yields a conservative bound $v_\Delta > \unit[10]{eV}$ in the preferred parameter space of Fig.~\ref{fig:ST}. A similar bound was derived in Ref.~\cite{Cheng:2022jyi}.
$\mu\to 3e$~\cite{SINDRUM:1987nra} naively gives stronger constraints on $v_\Delta$ up to keV, but only assuming that all entries in $M_\nu$ are of similar order; if nature has picked e.g.~$|(M_\nu)_{ee}|\ll |(M_\nu)_{\mu e}|$ instead -- thus suppressing neutrinoless double-beta decay beyond testability -- $\mu\to 3e$ no longer provides useful constraints on the triplet model. For these special regions of parameter space, lepton-flavor-violating $\tau$ decays then provide constraints of similar order to $\mu\to e\gamma$, namely $v_\Delta \gtrsim \unit[10]{eV}$, but depending on all unknown neutrino parameters.
Given all of these constraints, we will conservatively assume $v_\Delta > \unit{keV}$ and $|Y_{\alpha\beta}|<10^{-3}$ in the following, which ensures compliance with current rare-decay constraints in the mass range of interest.

Independent constraints on $Y$ arise from collider searches. The doubly-charged scalar $H^{++}$ is light enough to be pair-produced efficiently via the Drell--Yan process, and decays into two same-sign leptons $H^{++}\to \ell^+_\alpha \ell^+_\beta$ with rate $\Gamma \propto |Y_{\alpha\beta}|^2$~\cite{Melfo:2011nx}. If these are the dominant decay modes of $H^{++}$, then the entire preferred $1\sigma$ region of Fig.~\ref{fig:ST} is excluded by constraints on these channels~\cite{ATLAS:2014kca,ATLAS:2017xqs}. 
The only scenario allowing for $m_{H^{++}}<\unit[350]{GeV}$ is $Y\ll 1$ in order to suppress both lepton flavor violation \emph{and} the leptonic decay channels of $H^{++}$. Ultimately, we therefore require
\begin{align}
\unit{MeV} < v_\Delta < \unit{GeV}
\end{align}
to accommodate the new $W$ mass within the triplet seesaw model. With $Y\ll 1$, the dominant decay channels of $H^{++}$ become~\cite{Kang:2014lwn}
\begin{align}
\Gamma (H^{++}\to W^+ W^+)&\simeq \frac{g^4 v_\Delta^2 m_{H^{++}}^3}{64\pi m_W^4}
\end{align}
as well as $H^{++}\to W^+ (W^+\to \bar f f')$ in the region $m_W < m_{H^{++}}< 2 m_W$.
ATLAS~\cite{ATLAS:2018ceg,ATLAS:2021jol} searches exclude the $H^{++}$ mass region between $200$ and $\unit[350]{GeV}$ in the $pp\to H^{++}H^{--}\to 4W$ channel; see also Ref.~\cite{Ashanujjaman:2021txz}. Furthermore, a re-interpretation of ATLAS data gives $m_{H^{++}} >\unit[84]{GeV}$~\cite{Kanemura:2014goa,Kanemura:2014ipa} from the $pp\to H^{++}H^{--}\to 4W^*$ channel, as indicated in Fig.~\ref{fig:ST} (bottom).
In the remaining mass window $\unit[84]{GeV}\lesssim m_{H^{++}}\lesssim \unit[200]{GeV}$, the production cross section $pp\to H^{++}H^{--}$ is between $\unit[1]{pb}$ and $\unit[0.1]{pb}$ at the LHC~\cite{Fuks:2019clu} and  can be conclusively tested, as pointed out already in Ref.~\cite{Kang:2014lwn}. In parts of this preferred region the $H^{++}$ could even lead to displaced vertices~\cite{BhupalDev:2018tox,Antusch:2018svb}, since all decay rates of $H^{++}$ are suppressed by $v_\Delta^2$ or $|Y|^2$.
The triplet-seesaw explanation of the CDF result is hence completely testable through $H^{++}$; a \emph{confirmation} of this explanation requires also observing the neutral and singly charged partners though.

In the remaining mass window for the \emph{singly} charged scalar, $H^+$ is $60$--$\unit[80]{GeV}$ heavier than $H^{++}$ (Fig.~\ref{fig:ST}). The decays $H^+\to H^{++}W^-$ are then kinematically forbidden and only occur through off-shell $H^{++}$ or $W^-$, leading to multi-body signatures. The decay rate is approximately~\cite{Ashanujjaman:2021txz}
\begin{align}
\Gamma(H^+\to H^{++} {W^{-}}^*)\simeq \frac{3 m_{H^+}^5}{160\pi^3 v^4}\left( 1-\frac{m_{H^{++}}^2}{m_{H^+}^2}\right)^5 .
\label{eq:Hp_decay}
\end{align}
Notice that $H^+$ has couplings to leptons and quarks even in the limit $Y\ll 1$, albeit suppressed by $v_\Delta^2$ and hence unfortunately subdominant over the entire parameter space of interest.
At the higher mass end, $H^+\to h W^+$ could be sizable, but depends on free parameters in the scalar potential that couple the Higgs $h$ to the triplet.
Overall, Eq.~\eqref{eq:Hp_decay} is the most promising decay channel, which ultimately yields $H^+\to 3 W$ with at least one of the $W$ bosons being off-shell.

Finally, the neutral $H$ and $A$ bosons are predicted to lie between $200$ and $\unit[300]{GeV}$ (Fig.~\ref{fig:ST}), with a mass splitting of $\unit[50]{GeV}$ with respect to $H^+$ that is again too small to decay into on-shell $H^+ W^-$. The neutral scalars then dominantly decay through $H^+$ and $H^{++}$ with emission of up to four off-shell $W$ bosons. Other decay modes are suppressed by $v_\Delta$ or depend on free parameters~\cite{Ashanujjaman:2021txz}. While hardly the best signature, a search for $H$/$A$ is paramount should a $H^{++}$ be found in the predicted mass range.

CDF's new precise $W$-boson mass measurement differs from the SM expectation by a stunning seven standard deviations. This strongly implies a breakdown of custodial symmetry and can be explained with the help of new electroweak multiplets. One well-motivated example for a new multiplet comes in the form of a triplet with hypercharge, which has been discussed for many years as one of the simplest origins for Majorana neutrino masses. As shown here, this simple model can furthermore explain the CDF result, predicting doubly-charged scalars with mass between $100$ and $\unit[200]{GeV}$, completely testable at the LHC through Drell--Yan pair production and decays $H^{++}\to W^+ W^+$ and $H^{++}\to W^+ \bar f f'$. Singly-charged and neutral scalars with slighly heavier masses are predicted as well, albeit more difficult to detect. We hope that the CDF anomaly provides sufficient motivation to search for charged scalars in the \unit[100]{GeV} mass range in ATLAS and CMS.

\section*{Acknowledgements}
I thank Anil Thapa for discussions.

\bibliographystyle{utcaps_mod}
\bibliography{BIB}

\end{document}